\documentstyle[12pt]{article}
\begin{document}
\def\be{\begin{equation} }
\def\ee{\end{equation} }
\begin{titlepage}
\begin{flushright}
BRU/PH/205 \\
arch-ive/9510011 \\
October 1995 \\
\end{flushright}
\vskip4.5cm
\begin{center}
{\LARGE\bf On the  Kinematic Re\-con\-struc\-tion  }   \\
\vskip0.5cm
{\LARGE\bf of $e^+e^-\rightarrow W^+W^-\rightarrow jj\tau\overline{\nu}_\tau$\
Events}
\end{center}
\vspace{0.9cm}
\begin{center}
R. Rylko
\end{center}
\vskip0.2cm
\begin{center}
\footnotesize
Department of Physics, Brunel University, Uxbridge UB8 3PH, UK.
\end{center}
\vskip1cm
\centerline{ {\bf Abstract} }
\begin{quote}
{\small\rm\hspace*{6mm}
We show that the kinematic reconstruction of the
$e^+e^-\rightarrow W^+W^-\rightarrow jj\tau^-\overline{\nu}_\tau$\
events have a one-parameter ambiguity when reconstructed from the momentum of
all measured $W^-$\ decay products.
We propose a {\em hybrid} method of reconstruction of the
$e^+e^-\rightarrow W^+W^-\rightarrow jj\tau^-\overline{\nu}_\tau$\
events.
This is based on the observation that the difference between the $\tau^-$\
production angles
and the production angles of the sum of its visible decay products is small,
whilst the $\tau^-$\ energy  is poorly reconstructed.
This method consists of taking the $\tau^-$\ production angles from
those measured for the sum of the visible $\tau^-$\ decay products and
reconstructing the $\tau^-$\ energy from energy-momentum conservation
constraints. A reconstruction using this method is found to be well-defined and
possess a unique solution for the $\tau^-$\ momentum range at LEP II
and NLC.}
\end{quote}
\vskip2.0cm
{\em (Extended version of the contribution to the Report of Three Gauge
Couplings Working Group at the LEP II Workshop)}
\end{titlepage}
\setcounter{footnote}{0}
One of the subjects of  LEP II is to study
the three gauge couplings (TGC) of the $\gamma/Z$\ and $W$\
bosons. Due to expected limited statistics at LEP II,
one of the crucial points will be to include
as many $W$\ decay channels as possible for these studies.
Ongoing Monte Carlo studies use semileptonic
$WW\rightarrow jjl\overline{\nu}_l$\ (with $l$=$e$\ or $\mu$)
and $WW\rightarrow 4j$\ events.
The four jet channel, although the most abundant, is likely to
be affected
by the difficulties in jet tagging.
Thus, folded angular distributions will have to be used, which result in
a significant increase of the error for the fitted couplings
\cite{Sekulin}.

It is widely believed that $WW\rightarrow jj\tau\overline{\nu}_\tau$\
events may be used with the other semileptonic
$WW\rightarrow jjl\overline{\nu}_l$\ (with $l$=$e$\ or $\mu$) events, to
increase statistics for studies of the
triple gauge couplings
and/or the $W$\ boson mass measurements. Indeed,
for the semileptonic $e^+e^-\rightarrow
W^+W^-\rightarrow jj\tau^-\overline{\nu}_\tau$\ events
there are six unmeasured variables, namely the momentum of the
anti-neutrino  from $W^-$\ decay, ${\bf p}_{\overline{\nu}}^{lab}$,
and the momentum of neutrino from $\tau^-$\ decay, ${\bf p}_{\nu}^{lab}$.
There are also six constraints, namely (in the narrow $W$\ width
approximation):
\be
{\bf p}_{vis}^{lab} + {\bf p}_{\nu}^{lab}
   + {\bf p}_{\overline{\nu}}^{lab} = {\bf p}_W^{lab}
\ee
\be
E_{vis}^{lab} + p_{\nu}^{lab} +  p_{\overline{\nu}}^{lab}
          + E_W^{lab} = \sqrt{s}
\ee
\be
(E_{vis}^{lab} + p_{\nu}^{lab} + p_{\overline{\nu}}^{lab} )^2
   - ({\bf p}_{vis}^{lab} +  {\bf p}_{\nu}^{lab}
   + {\bf p}_{\overline{\nu}}^{lab})^2
                                       = m_W^2
\ee
\be
(E_{vis}^{lab} + p_{\nu}^{lab})^2
      - ({\bf p}_{vis}^{lab} + {\bf p}_{\nu}^{lab})^2 = m_{\tau}^2
\ee
where $E_{vis}^{lab}$\ and ${\bf p}_{vis}^{lab}$\ are the energy and
momentum of the visible (i.e.\ charged and neutral $\tau^-$\ decay products
other than neutrino) $\tau^-$\
decay products respectively,
$E_W^{lab}$\ and ${\bf p}_W^{lab}$\ are the energy and momentum of
the other boson ($W^+$)
reconstructed from the two jets respectively,
$\sqrt{s}$\ is the $e^+e^-$\ center of mass
energy and $m_{\tau}$\ is the $\tau$\ lepton mass.
The above equations can be reduced to equations for the energies
of the anti-neutrino and
neutrino and three angles.
Thus, it is difficult to see whether or not there is an ambiguity in the above
equations, and what the character of this ambiguity might be.
As the above equations contain quadratic terms of
momenta, one may expect a two-fold ambiguity.

The problem simplifies when solved in the $W^-$\ rest frame. The boost to this
frame is uniquely defined by the momentum of the $W^+$\ boson
determined from the measurement of the two jets.
The above  set of constraints now reduces to the following equations:
\be     \label{plane}
{\bf p}_{vis} + {\bf p}_{\nu} + {\bf p}_{\overline{\nu}} = 0
\ee
\be
E_{vis} +  p_{\nu} +  p_{\overline{\nu}}= m_W
\ee
\be
(E_{vis} + p_{\nu})^2 - ({\bf p}_{vis} + {\bf p}_{\nu})^2 = m_{\tau}^2 \;\; .
\ee
In the $W^-$\ rest frame all three momenta  lie
in a plane.
The boosted visible $\tau$\ decay products' momentum
${\bf p}_{vis}$\ fixes the plane defined by eq.\ (\ref{plane}), but there is
a freedom of rotation of the plane around ${\bf p}_{vis}$.
The solution of these equations is:
\be
p_{\nu}  =  \frac{m_W^2 - 2 m_W E_{vis} + m_\tau^2} {2 m_W}
\ee
and
\be
p_{\overline{\nu}}  =  \frac{m_W^2 - m_\tau^2} {2 m_W} \;\; ,
\ee
and the cosines of angles between  ${\bf p}_{vis}$,
${\bf p}_{\nu}$\ and ${\bf p}_{\overline{\nu}}$\
may be expressed
in terms of $p_{\nu}$\  and  $p_{\overline{\nu}}$.
The other solutions may be obtained from the one above by rotations
of the decay plane around ${\bf p}_{vis}$.
Thus, the $W^-$\ helicity
angle $\theta^*$, i.e.\ the angle between $W^-$\ laboratory momentum and
the momentum of $\tau$\ in the $W^-$\ rest frame, as well as
the helicity angle $\phi^*$, lie
in a certain range resulting from the full rotation
of ${\bf p}_{\nu}$\ around ${\bf p}_{vis}$.
This ambiguity in the reconstruction of the $\tau$\ momentum
(${\bf p}_\tau = {\bf p}_{vis} + {\bf p}_\nu$)
in the $W^-$\ momentum rest frame, when boosted
back to the laboratory frame, leads to ambiguous energy and production
angles for $\tau^-$.
Thus, the events may not be fully reconstructed kinematically, i.e.\
there is a one parameter ambiguity for a kinematic fit similar
to the one for the $W^+W^-\rightarrow jjl\overline{\nu}_l$\
(with $l$=$e$\ or $\mu$) case.

One may consider the $\tau$\ reconstruction method based on finding its
flight vector from the secondary vertex.
Although the momentum, and thus the decay length of the
$\tau$\ is large, however, due to small branching ratio for the $\tau$\
decaying to more than one charged particle,
the momentum estimators (see e.g.\ \cite{RR} and references
cited therein) seem unlikely to be applied successfully.
On the other hand, at LEP II energies, the $\tau$\ lepton
from $e^+e^-\rightarrow W^+W^-\rightarrow jj\tau^-\overline{\nu}_\tau$\
will have high momentum compared with
the transverse (to the direction of $\tau$) momenta of its decay
products, which are of the order of $m_\tau$/2.
Thus, one may expect
that laboratory $\tau$\ production angles are almost the same as
the production angles of the sum of its visible decay products.
Indeed, the simulation using PYTHIA 5.7/JETSET 7.4 \cite{PYTJET}
at $\sqrt{s}$=175 GeV, shows that
for 88\%\ of all hadronic $\tau$\ decays the difference between the cosine
of the $\tau$\ polar production angle and the cosine of the polar production
angle of the sum of its visible products is within $\pm$0.1.
Also for 91\%\ of the decays
this difference for the azimuthal production angle is within $\pm$0.2 rad.
The energy of the $\tau$\ is, however, substantially different than
the energy of its visible decay products.

This suggests two methods. The first method is to study the distribution
of the production angles
only (assuming the angles of ${\bf p}_{vis}^{lab}$\
to be the $\tau$\ production
angles), resulting in a fit with less angular information, and thus
a greater degree of inaccuracy.
The other {\em hybrid} method is to use the measured production angles of
the visible $\tau$\ decay products as
the $\tau$\ production angles and reconstruct the $\tau$\ energy.
Let $\hat{\tau}$\ be the unit vector in the
direction of the momentum of the sum of the visible $\tau^-$ decay products.
Now, assuming that $\hat{\tau}$\ is
the true $\tau^-$\ direction
gives the following constraints for the laboratory $\tau^-$\  momentum,
$p_\tau^{lab}$, and
the neutrino laboratory momentum vector, ${\bf p}_\nu^{lab}$:
\be    \label{CONSTR1}
 p_\tau^{lab} \hat{\tau} +  {\bf p}_\nu^{lab} = - {\bf p}_W^{lab}
\ee
and
\be     \label{CONSTR2}
 E_\tau^{lab}  +   p_\nu^{lab}  =  \sqrt{s} - E_W^{lab} \;\; ,
\ee
where $E_W^{lab}$\ and ${\bf p}_W^{lab}$\ are the laboratory energy and
momentum of the
other $W^+$\ boson reconstructed from the $jj$\ final state, respectively.
The above four equations may be reduced to one quadratic equation.
For  $p_W^{lab}<(m_{W^-}^2 - m_\tau^2)/2 m_\tau$, i.e.\ for
$p_W^{lab}<(1.8\pm 0.1)$\ TeV for nominal $W$\ boson mass \cite{PDG} (where
a change in the nominal $W$\ mass within one
width produces this variation of 0.1 TeV),
there are two solutions of the above equations, namely:
\be     \label{wynik}
p_{\tau \pm}^{lab} = \frac{ - p_W^{lab}
                      \cos\theta_{\tau^- W^+}\Sigma^2  \pm
                  (\sqrt{s}-E_W^{lab}) \sqrt{\Sigma^4 - 4m_\tau^2
                                 [(\sqrt{s}-E_W^{lab})^2 - p_W^{lab\; 2}
                                           \cos^2\theta_{\tau^- W^+}] }
                     }
               { 2 [(\sqrt{s}-E_W^{lab})^2 - p_W^{lab\; 2}
                                \cos^2\theta_{\tau^- W^+}] }
\ee
where $\theta_{\tau^- W^+}$\ is the angle between $\hat{\tau}$\ and the
other $W^+$\ momentum, and
\be
\Sigma^2 = (\sqrt{s}-E_W^{lab})^2 - p_W^2 + m_\tau^2   \;\; .
\ee
However, for the $W^-$\ boson energies
\be
E_{W^-}^{lab}    <  \frac{m_{W^-}^2 + m_\tau^2}{2 m_\tau}  \;\; ,
\ee
one of the above solutions is negative and the other is positive,
thus there is only one physical solution to the above problem.

The above property shows that the  proposed {\em hybrid} method may be used
for kinematic fitting, where the $\tau$\ production angles
are fitted to the measured values of the production
angles of the sum of the visible $\tau$\ decay products,
and the $\tau$\ momentum value $p_\tau^{lab}$\ is fitted using the
constraints given by eq.(\ref{CONSTR1}) and eq.(\ref{CONSTR2}).
As a narrow $W$\ width approximation is not implied in eq.(\ref{CONSTR1})
and eq.(\ref{CONSTR2}),
the above method may be used to exploit the
$e^+e^-\rightarrow W^+W^-\rightarrow jj\tau\overline{\nu}_\tau$\
for TGC studies as well as for
$W$\ boson mass measurements.

\vskip0.1cm
\begin{flushleft}
{\large\bf Acknowledgements}
\vskip0.1cm
\end{flushleft}
\hspace*{6mm} The author thanks P.Clarke, J.Conboy, A.Hasan, R.Sekulin,
P.Singh and A.Skillman for remarks.


\vskip0.1cm
\begin{thebibliography}{99}
\bibitem{Sekulin} R.L.Sekulin, Phys. Lett. {\bf B338}(1994)369.
\bibitem{RR}  R.Rybicki, R.Rylko,   Phys. Lett. {\bf B353}(1995)547.
\bibitem{PYTJET} T.Sj\"{o}strand, PYTHIA 5.7 and JETSET 7.4, Physics and
Manual, \\ CERN-TH.7112/93.
\bibitem{PDG} Particle Data Group, L.Montanet et al., Phys. Rev.
{\bf D50}(1994)1173.
\end{thebibliography}
\end{document}